\documentclass[runningheads]{llncs}
\usepackage[T1]{fontenc}
\usepackage{graphicx}
\usepackage{microtype}
\usepackage{booktabs}
\usepackage{array}
\usepackage{amsfonts}
\usepackage{amsmath}
\usepackage{amssymb}
\usepackage{multirow}
\usepackage{multicol}
\usepackage{listings}
\usepackage{placeins}
\usepackage{subfigure}
\usepackage{graphicx}
\usepackage{hyperref}
\usepackage[misc]{ifsym}

\usepackage{color}

\begin{document}
\title{Probing Spurious Correlations in Popular Event-Based Rumor Detection Benchmarks}
%
%
\author{Jiaying Wu\textsuperscript{(\Letter)} \and
Bryan Hooi}
\authorrunning{J. Wu and B. Hooi}
\tocauthor{Jiaying Wu and Bryan Hooi}

\titlerunning{Spurious Correlations in Popular Event-Based Rumor Detection Benchmarks}
\toctitle{Probing Spurious Correlations in Popular Event-Based Rumor Detection Benchmarks}
%
\institute{School of Computing, National University of Singapore, Singapore, Singapore
\email{jiayingwu@u.nus.edu, bhooi@comp.nus.edu.sg}}

\maketitle              
\begin{abstract}
As social media becomes a hotbed for the spread of misinformation, the crucial task of rumor detection has witnessed promising advances fostered by open-source benchmark datasets. Despite being widely used, we find that these datasets suffer from spurious correlations, which are ignored by existing studies and lead to severe overestimation of existing rumor detection performance. The spurious correlations stem from three causes: (1) \emph{event-based} data collection and labeling schemes assign the same veracity label to multiple highly similar posts from the same underlying event; (2) merging multiple data sources spuriously relates source identities to veracity labels; and (3) labeling bias. In this paper, we closely investigate three of the most popular rumor detection benchmark datasets (i.e., Twitter15, Twitter16 and PHEME), and propose \emph{event-separated rumor detection} to eliminate spurious cues. Under the event-separated setting, we observe that the accuracy of existing state-of-the-art models drops significantly by over 40\%, becoming only comparable to a simple neural classifier. To better address this task, we propose Publisher Style Aggregation (PSA), a generalizable approach that aggregates publisher posting records to learn writing style and veracity stance. Extensive experiments demonstrate that our method outperforms existing baselines in terms of effectiveness, efficiency and generalizability. 

\end{abstract}
\section{Introduction}
In the battle against escalating online misinformation, recent years have witnessed growing interest in automatic rumor detection on social media. Numerous real-world rumor detection datasets including \textsc{Twitter15} \cite{ma17detect}, \textsc{Twitter16} \cite{ma17detect} and \textsc{PHEME} \cite{kochkina18one} have emerged as valuable resources fueling continuous development in this field. From early feature engineering models \cite{castillo11information} to recent content-based \cite{zhang19fakedetector,ma18rumor} and propagation-based \cite{bian20rumor,huang20conquering} methods, the ever-evolving approaches have achieved promising advances.

However, most existing methods ignore the \textit{spurious attribute-label correlations} induced by dataset construction pitfalls, which arise from dataset-related artifacts instead of relationships generalizable to practical real-world settings. The commonly adopted \textit{event-based} data collection framework first fact-checks newsworthy events, and then automatically scrapes a large number of highly similar microblogs (e.g., tweets) containing the same event keywords. Some benchmark datasets also merge data samples from multiple existing sources to balance their class distribution. These factors consequently induce event-label and source-label correlations, which may not hold in practice. 

Negligence of spurious cues can lead to unfair over-predictions that limit model generalization and adaptability. Similar issues have been identified in several natural language processing tasks including sentiment classification \cite{wang20identifying}, argument reasoning comprehension \cite{niven19probing} and fact verification \cite{schuster19towards}, but the task of social media rumor detection remains underexplored.

\begin{figure}[t] 
 \centerline{\includegraphics[width=\textwidth]{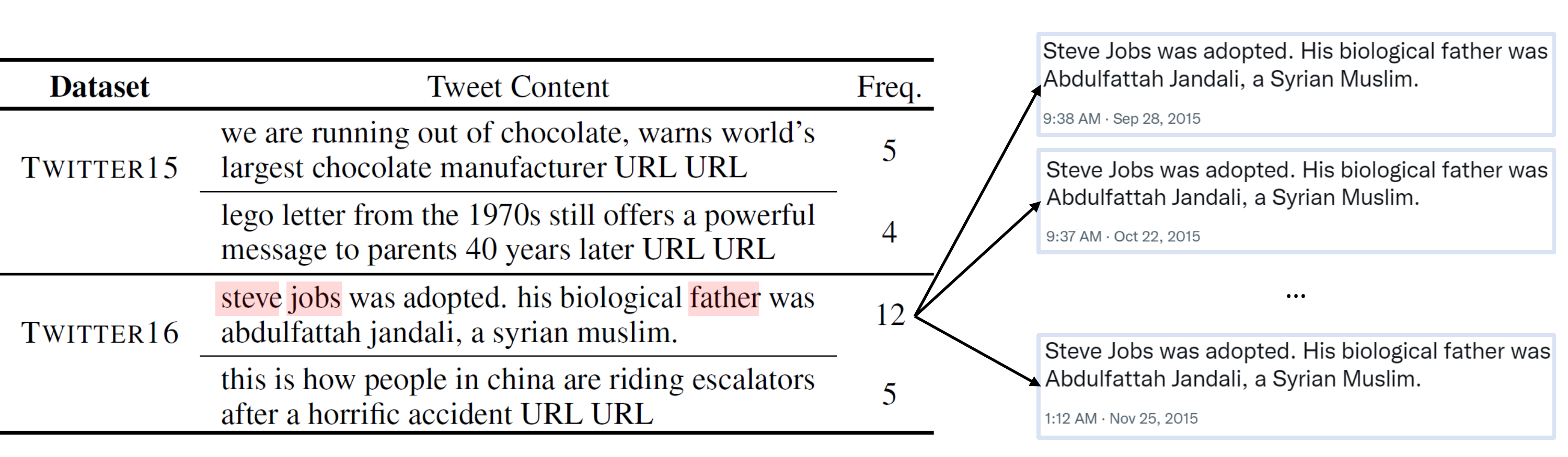}}
 \caption{Automated event-based scraping results in numerous duplicate microblog texts in benchmark datasets \textsc{Twitter15} and \textsc{Twitter16}, causing data leakage under random splitting. The highlighted words are event keywords obtained from the Snopes fact-checking website, in line with the datasets' data collection scheme (see Section \ref{data-collect}).}
 \label{fig:text-dup}
\end{figure}

Hence, in this paper, we make an effort to thoroughly investigate the causes of spurious correlations in existing rumor detection benchmark datasets, and take solid steps to counteract their impact. Specifically, we identify three causes of spurious correlations: (1) event-based data collection and labeling strategies associate event keywords with veracity labels; (2) merging data sources for label balancing creates correlations between source-specific propagation patterns and microblog veracity; and (3) event-level annotation strategies give rise to labeling bias. Under the post-level data splitting scheme commonly adopted by existing approaches, the prevalence of spurious cues can bring about numerous shared spurious correlations between the training and test data. For instance, the training and test data might even contain identical microblog texts, leading to data leakage (see examples in Fig. \ref{fig:text-dup}). If left unchecked, these correlations can lead to severe overestimation of model performance.

To offset the impact of spurious cues, we study a more practical task, namely \textit{event-separated rumor detection}, where the test data contains microblogs from a set of events unseen during training. Without prior knowledge of event-specific cues in the test set, we empirically demonstrate stark performance deterioration of existing approaches, e.g. state-of-the-art rumor detection accuracy plummets from 90.2\% to 44.3\% on \textsc{Twitter16} \cite{ma17detect}, one of the most widely adopted datasets. 

Striving for reliable rumor detection, we propose Publisher Style Aggregation (PSA), a novel method inspired by human fact-checking logic (i.e. reading through a user's homepage to determine user stance and credibility). Specifically, our approach (1) encodes the textual features of source posts and user comments; (2) learns publisher-specific features based on multiple microblog instances produced by each source post publisher; and (3) augments each local microblog representation with its corresponding global publisher representation. 

We evaluate PSA on the event-separated rumor detection task using 3 real-world benchmark datasets and compare it against 8 state-of-the-art baselines. Extensive experiments show that PSA outperforms its best competitors by a significant margin, respectively boosting test accuracy and F1 score by 14.18\% and 15.26\% on average across all 3 datasets. Furthermore, we empirically demonstrate the efficiency and generalizability of PSA via two experimental objectives, namely early rumor detection and cross-dataset rumor detection. Our code is publicly available at: \url{https://github.com/jiayingwu19/PSA}.

\section{Related Work}

\noindent\textbf{Social Media Rumor Detection} 
Real-world rumor detection datasets, with microblog posts and propagation patterns retrieved from social media platforms such as Twitter \cite{liu15realtime,ma17detect,kochkina18one} and Weibo \cite{ma16detecting}, form the bedrock of rapidly evolving approaches. 

Recent advances in automated rumor detection typically adopt neural network based frameworks. \textit{Content-based} approaches utilize microblog and comment features. For instance, \cite{ma16detecting,liu18early} respectively employ Recurrent Neural Networks and Convolutional Neural Networks to model the variations of text and user representations over time. Hierarchical attention networks \cite{shu19defend} and pre-trained language models \cite{pelrine21surprising} have also proven effective. Another line of work leverages \textit{propagation-based} information diffusion patterns to encode information flow along user interaction edges. Some inject structural awareness into recursive neural networks \cite{ma18rumor} and multi-head attention \cite{yuan19jointly,yuan20early,khoo20interpret}, while others achieve success with Graph Neural Networks \cite{bian20rumor,huang20heterogeneous}. 

Closely related to our topic is fake news detection. \cite{wang18eann} trains an event discriminator to overlook domain-specific knowledge under the multi-modal setting, \cite{han20graph} formulates domain-agnostic fake news detection as a continual learning problem, \cite{silva21embracing} studies the case with limited labeling budget, and \cite{nguyen20fang,dou21user} take advantage of auxiliary user descriptions and large-scale user corpus.

Existing methods either overlook the publisher-microblog relationship or require external knowledge (e.g. images and additional user description). In contrast, we seek to achieve generalizable rumor detection by capturing rumor-indicative publisher characteristics based on aggregation of multiple microblog data samples.
 
\noindent\textbf{Investigation of Spurious Correlations} 
Despite the promising performance of deep learning models, reliance on dataset-related cues has been observed in a wide range of tasks including text classification \cite{wang20identifying}, natural language inference \cite{mccoy19right} and visual question answering \cite{agarwal20towards}. In fact-checking scenarios, language models can capture underlying identities of news sites \cite{zhou21hidden}, and rumor instances can possess time-sensitive characteristics \cite{pelrine21surprising}. Spurious artifacts lead to model failure on out-of-domain test instances, as empirically observed by \cite{khoo20interpret,huang20conquering,sagawa20inv}.

However, systematic investigation into social media rumor detection remains unexplored. We bridge this gap by discussing three types of spurious correlations specific to this topic, and provide a solution to offset the impact of spurious correlations (i.e. event-separated rumor detection).

\section{Spurious Correlations in Event-Based Datasets} \label{spurious-dset}

\subsection{Event-Based Data Collection} \label{data-collect}

In this subsection, we outline the event-based data collection scheme adopted by benchmark datasets.

\textbf{Newsworthy Event Selection:} Newsworthy events serve as vital information sources, from which rumors and non-rumors arise and diffuse on social media. Existing studies either collect events from leading fact-checking websites (e.g., Snopes, Emergent, and PolitiFact) \cite{ma16detecting,liu15realtime,shu18fakenewsnet}, or obtain candidate events identified by professionals \cite{kochkina18one}.

\textbf{Keyword-Based Microblog Retrieval:} To facilititate mass collection, existing datasets typically adopt automated \textit{event-based} data collection strategies, i.e. for each event, (1) extract keywords from its claim; (2) scrape microblogs via keyword-based search; and (3) select influential microblogs. Event keywords are mostly \textit{neutral} (e.g. places, people or objects), carrying little or no stance.

\textbf{Microblog Labeling Scheme:} Existing rumor detection datasets conduct fact-checking at either event-level \cite{ma16detecting,liu15realtime,shu18fakenewsnet} or post-level \cite{kochkina18one}. While \textit{event-level} labeling assigns all source posts under an event with the same event-level fact-checking label, \textit{post-level} labeling annotates every source post independently. Although both event- and post-level annotations are performed by trained professionals, the former is more vulnerable to data selection pitfalls, on which we elaborate in Section \ref{spurious-cues}.

\subsection{Possible Causes of Spurious Correlations}\label{spurious-cues}

We investigate three of the most popular event-based rumor detection benchmark datasets containing source posts, propagation structures and conversation threads, namely \textsc{Twitter15} \cite{ma17detect}, \textsc{Twitter16} \cite{ma17detect} and \textsc{PHEME} \cite{kochkina18one}, and summarize the dataset statistics in Table \ref{tab:ds-stats}. As \textsc{Twitter15} and \textsc{Twitter16} both consist of class-balanced tweets with abundant interactions, we also sample class-balanced tweets involving at least 10 users in \textsc{PHEME} for a fair comparison.

\begin{table}[t]
\centering
\caption{Dataset statistics.} \label{tab:ds-stats}
 \begin{tabular}{lccc} \toprule
 \textbf{Dataset} &  \textsc{Twitter15} & \textsc{Twitter16} & \textsc{PHEME} \\ 
 \toprule
 Labeling Scheme & Event-Level  & Event-Level & Post-Level\\ 
 \# of Events & 298 & 182 & 9 \\
 \# of Source Posts & 1490 & 818 & 973 \\ 
 \# of Non-Rumors & 374 & 205 & 245 \\ 
 \# of False Rumors & 370 & 205 & 241 \\
 \# of True Rumors & 372 & 207 & 244 \\ 
 \# of Unverified Rumors & 374 & 201  & 243 \\ 
 \# of Distinct Users & 480,984 & 289,675 & 12,905\\ 
 \# of User Interactions & 622,927 & 362,713 & 21,169\\
 \bottomrule
\end{tabular}
\end{table}
\begin{figure}[t] 
 \centerline{\includegraphics[width=0.65\textwidth]{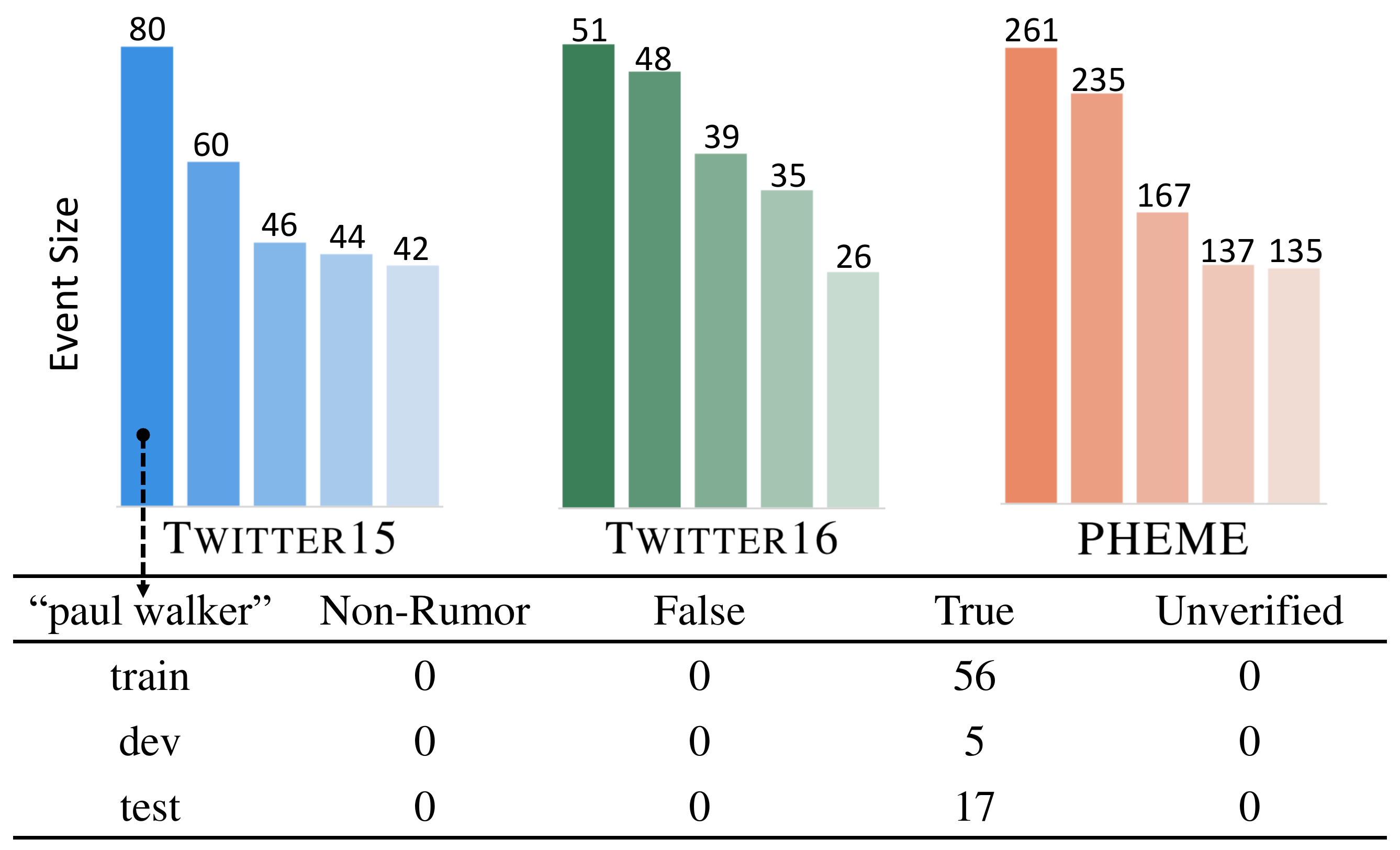}}
 \caption{The size of largest events in three datasets; ``paul walker'' directly correlates with ``True'' label in the post-level random splits adopted by the SOTA method \cite{yuan20early}.}
 \label{fig:event-size}
\end{figure} 

\noindent\underline{\textbf{Intra-Event Textual Similarity:}} 
Under each event, the automated keyword-based microblog retrieval framework collects a large number of highly similar keyword-sharing samples with the same label, even obtaining identical microblog texts (Fig. \ref{fig:text-dup}). Consequently, the correlations between event keywords and class labels result in strong textual cues that generalize poorly beyond the current event. Under the post-level data splitting scheme adopted by existing works, these cues would scatter into different splits, creating shared correlations between the training and test data that do not hold in the real world. We illustrate such correlations via \textsc{Twitter15}'s largest event about the death of Paul Walker (Fig. \ref{fig:event-size}). All 80 microblogs reporting this event are assigned the “True” label, among which 78 contain the keywords ``paul walker''. These event-specific keywords produce a strong correlation between ``paul walker'' and the ``True'' label. Under post-level random splitting, these samples spread across different data splits, creating undesirable textual similarity between the training and test data. As shown in Fig. \ref{fig:event-size}, the datasets are dominated by such large-size events. Specifically, the top-5 largest events cover 96.09\% of data samples in \textsc{PHEME}, while large-size events (containing more than 5 keyword-sharing tweets) cover more than 70\% of samples in \textsc{Twitter15} and \textsc{Twitter16}. Large event sizes lead to the prevalence of event-specific keyword-label correlations, further exacerbating the problem.

\begin{figure}[t] 
 \centerline{\includegraphics[width=0.8\textwidth]{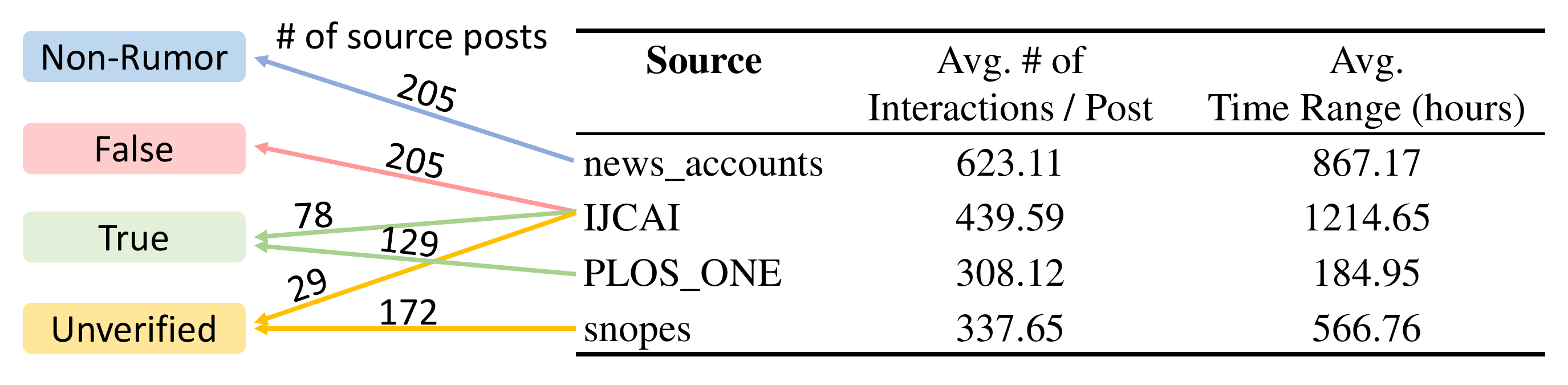}}
 \caption{Source-label correlations in \textsc{Twitter16} suggest underlying spurious properties. IJCAI and PLOS\_ONE refer to \cite{ma16detecting} and \cite{zubiaga16analysing}, respectively.}
 \label{fig:source-label-corr} 
\end{figure} 

\noindent\underline{\textbf{Merge of Data Sources:}}
For label balancing purposes, \textsc{Twitter15} and \textsc{Twitter16} merge tweets from multiple sources including \cite{ma16detecting,liu15realtime,zubiaga16analysing}, and scrape additional news events from verified media accounts. While the events covered by different sources do not overlap, direct 1-1 correlations between data sources and labels can possibly induce spurious correlations between data source features and the labels. As demonstrated in Fig. \ref{fig:source-label-corr}, the user interaction count (comments and reposts) and interaction time range of tweets from each source form distinctive patterns. For instance, all tweets from PLOS\_ONE \cite{zubiaga16analysing} are ``True'', spread very quickly and tend to arouse less interactions. These source-specific propagation patterns could possibly be exploited by graph- or temporal-based models, which we empirically demonstrate in Section \ref{event-insensitive} (Table \ref{tab:insensitive-cues}).

\noindent\underline{\textbf{Labeling Bias:}} 
\begin{figure}[t] 
 \centerline{\includegraphics[width=0.8\textwidth]{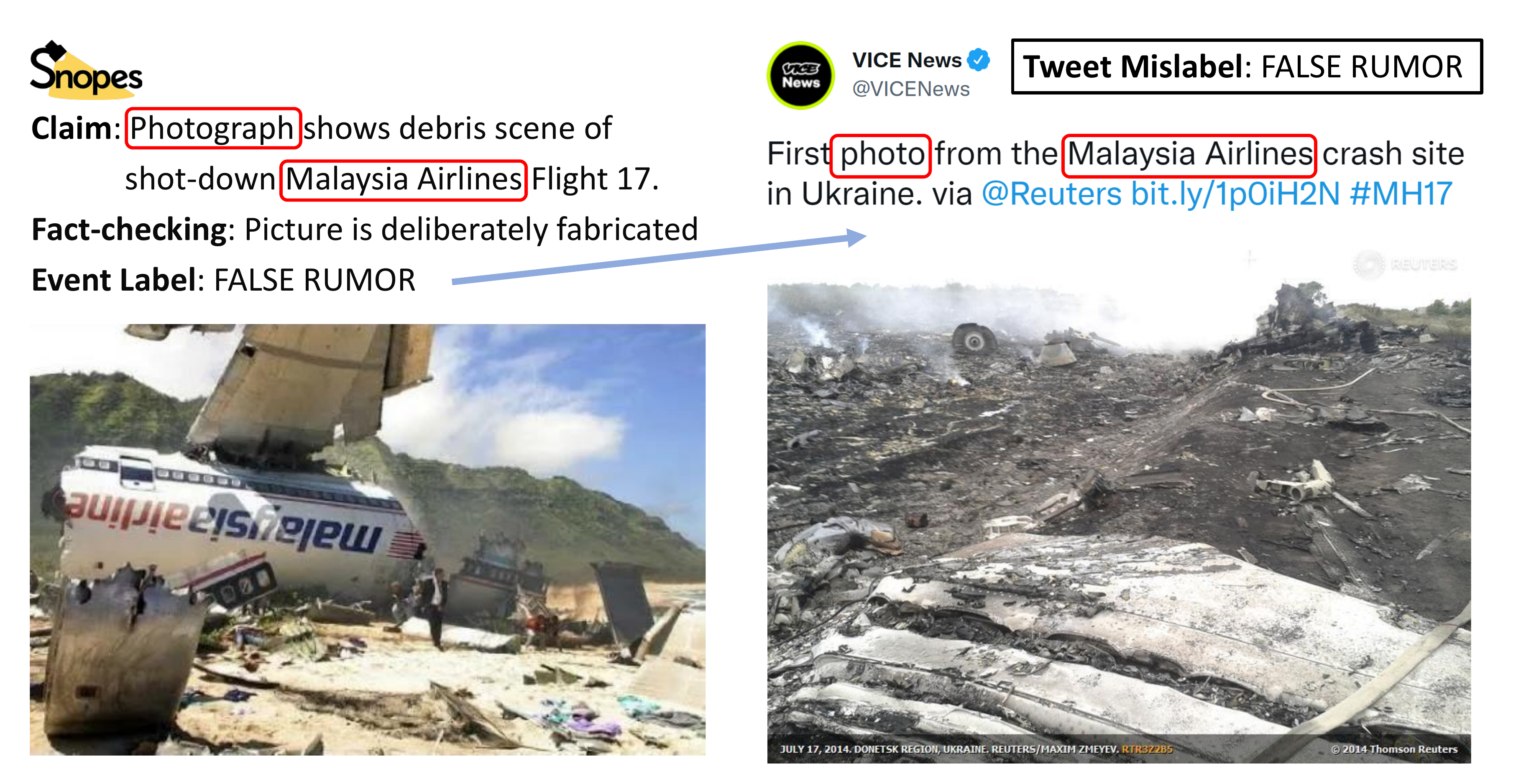}}
 \caption{One mislabeled instance from \textsc{Twitter15}, where the source tweet, inconsistent with the Snopes claim, is wrongly assigned the same event-level label.}
 \label{fig:mislabel-example} 
\end{figure} 
While automated event-based data retrieval and event-level annotations allow for easier construction of large-scale benchmark datasets, the lack of post-level scrutiny induces vulnerability to labeling bias. For instance, as shown in Fig. \ref{fig:mislabel-example}, Snopes marked a MH17-related event claim as ``False'' due to image fabrication. However, in view of highly similar keywords ``Malaysian Airlines'' and ``photo'', the data collection framework retrieved an MH17-relevant tweet linking to an authentic photo by Reuters and mistakenly labeled it as ``False'' in \textsc{Twitter15}. Such mislabelings exacerbate our previously mentioned problem of intra-event textual similarity, making the resulting keyword-label correlations stronger but more deceptive. To make the best use of valuable data resources, we suggest that future approaches incorporate techniques that are robust to label noise (e.g. noise-tolerant training \cite{li19learning}).

\section{Event-Separated Rumor Detection}
\label{sec:event-sep}

\subsection{Problem Formulation} \label{problem-form}

Social media rumor detection aims to learn a classification model that is able to detect and fact-check rumors. Let $\mathcal{T}=\{T_{1}, T_{2}, \ldots, T_{N}\}$ be a rumor detection dataset of size $N$, and $\mathcal{Y}=\{y_{1}, y_{2}, \ldots, y_{N}\}$ be the corresponding ground-truth labels, with $y_i \in \mathcal{C}=\{1,\dots, C\}$. Each microblog instance $T_{i}$ consists of a source post publisher $u_i$, a source post $p_{i}$, and related comments $c_{i}^{1}, \ldots, c_{i}^{k}$. 
$p_{i}$ ($c_{i}^{j}$) has a corresponding textual feature vector $\mathbf{r}_{i}$ ($\mathbf{r}_{i}^{j}$). 
Denote the event behind microblog instance $T_{i}$ as $e_i$. Consequently, training data $\mathcal{T}_{tr}$ and test data $\mathcal{T}_{te}$ respectively contain events $\mathcal{E}_{tr}=\left\{e_i| T_i\in \mathcal{T}_{tr}\right\}$ and $\mathcal{E}_{te}=\left\{e_i| T_i\in \mathcal{T}_{te}\right\}$. 

Most existing approaches ignore the underlying microblog-event relationship and adopt event-mixed post-level data splits, resulting in significant overlap between $\mathcal{E}_{tr}$ and $\mathcal{E}_{te}$. However, prior knowledge of test data is not always guaranteed in practice (e.g. the model's performance gains from duplicate tweets in the training and test data are unlikely to generalize), and previous assumptions can lead to performance overestimation caused by intra-event textual similarity (see Section \ref{spurious-cues}).

In order to eliminate these confounding event-specific correlations, we propose to study a more practical problem, namely \textbf{event-separated rumor detection}, where $\mathcal{E}_{tr} \cap \mathcal{E}_{te}=\varnothing$. This task is challenging due to the underlying event distribution shift, and it thereby provides a means to evaluate debiased rumor detection performance. 

\subsection{Existing Approaches} 
\label{existing-models}
We compare the event-mixed and event-separated rumor detection performance of representative approaches on \textsc{Twitter15}, \textsc{Twitter16} and \textsc{PHEME} datasets to investigate the impact of event-specific spurious correlations.

\noindent\underline{\textbf{Propagation-Based:}}\textbf{(1) TD-RvNN} \cite{ma18rumor}: a recursive neural model that encodes long-distance user interactions with gated recurrent units. \textbf{(2) GLAN} \cite{yuan19jointly}: a global-local attentive model based on microblog-user heterogeneous graphs. \textbf{(3) BiGCN} \cite{bian20rumor}: a GCN-based model that encodes augmented bi-directional rumor propagation patterns. \textbf{(4) SMAN} \cite{yuan20early}: a multi-head attention model that enhances training with user credibility modeling. 

\noindent\underline{\textbf{Content-Based:}} \textbf{(1) BERT} \cite{devlin19bert}: a deep language model that encodes bidirectional context with Transformer blocks. \textbf{(2) XLNet} \cite{yang19xlnet}: a generalized autoregressive approach that trains on all possible permutations of word factorizations. \textbf{(3) RoBERTa} \cite{liu19roberta}: a refined BERT-like approach that adopts dynamic masking for varied masking patterns over different epochs. \textbf{(4) DistilBERT} \cite{sanh20distilbert}: a distilled model that maintains 97\% of BERT's performance with 60\% less parameters.

\subsubsection{Data Splitting} For all three datasets, we sample 10\% instances for validation, then split the remainder 3:1 into training and test sets. Specifically, we obtain the event-separated splits based on the \textit{publicly available event IDs} released in \textsc{Twitter15} \cite{ma17detect}, \textsc{Twitter16} \cite{ma17detect} and \textsc{PHEME} \cite{kochkina18one}, respectively.

\begin{figure*}[t]
    \centering
    \subfigure[\textsc{Twitter15}]{\includegraphics[width=0.32\textwidth]{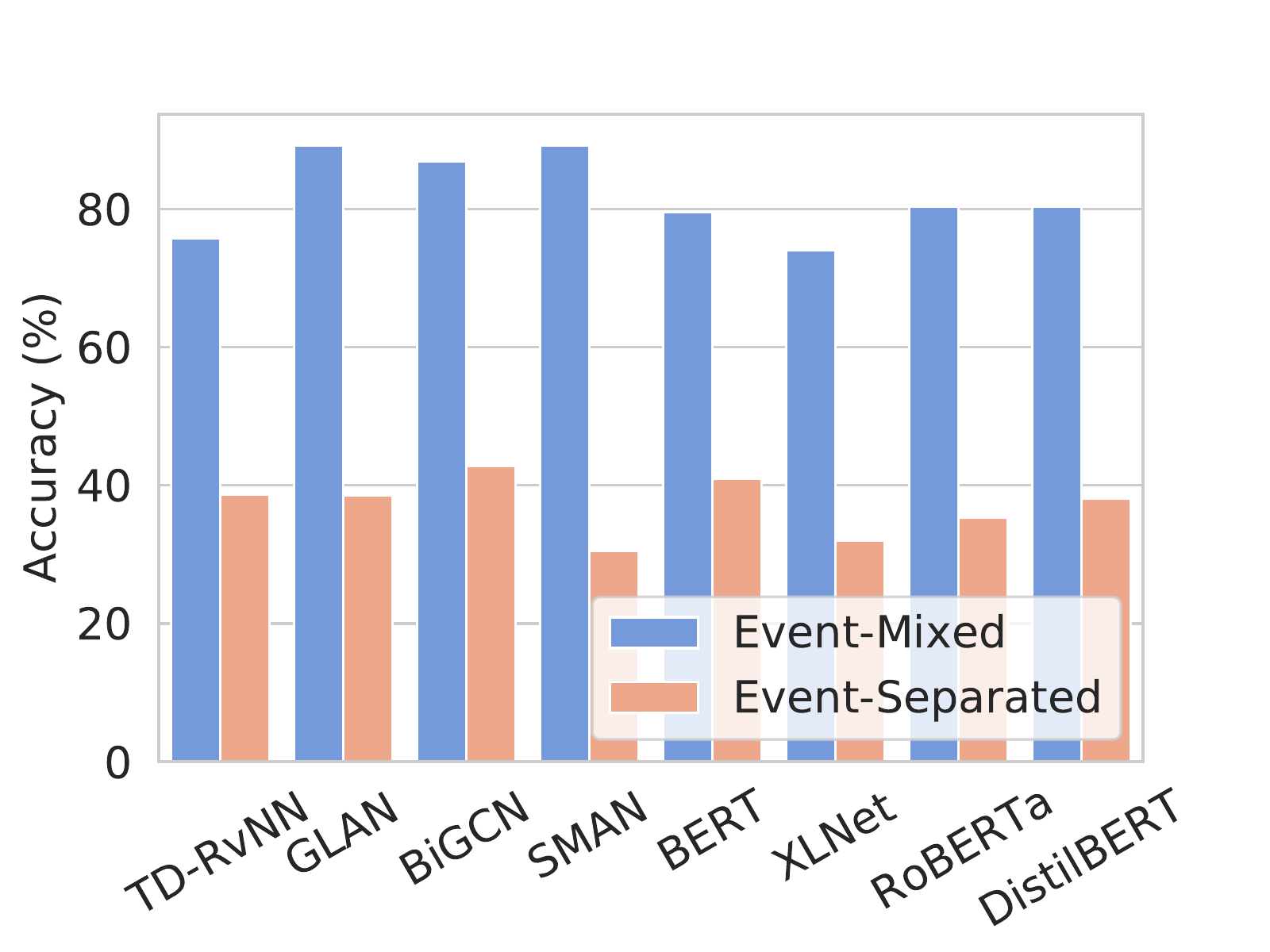}}
    \subfigure[\textsc{Twitter16}]{\includegraphics[width=0.32\textwidth]{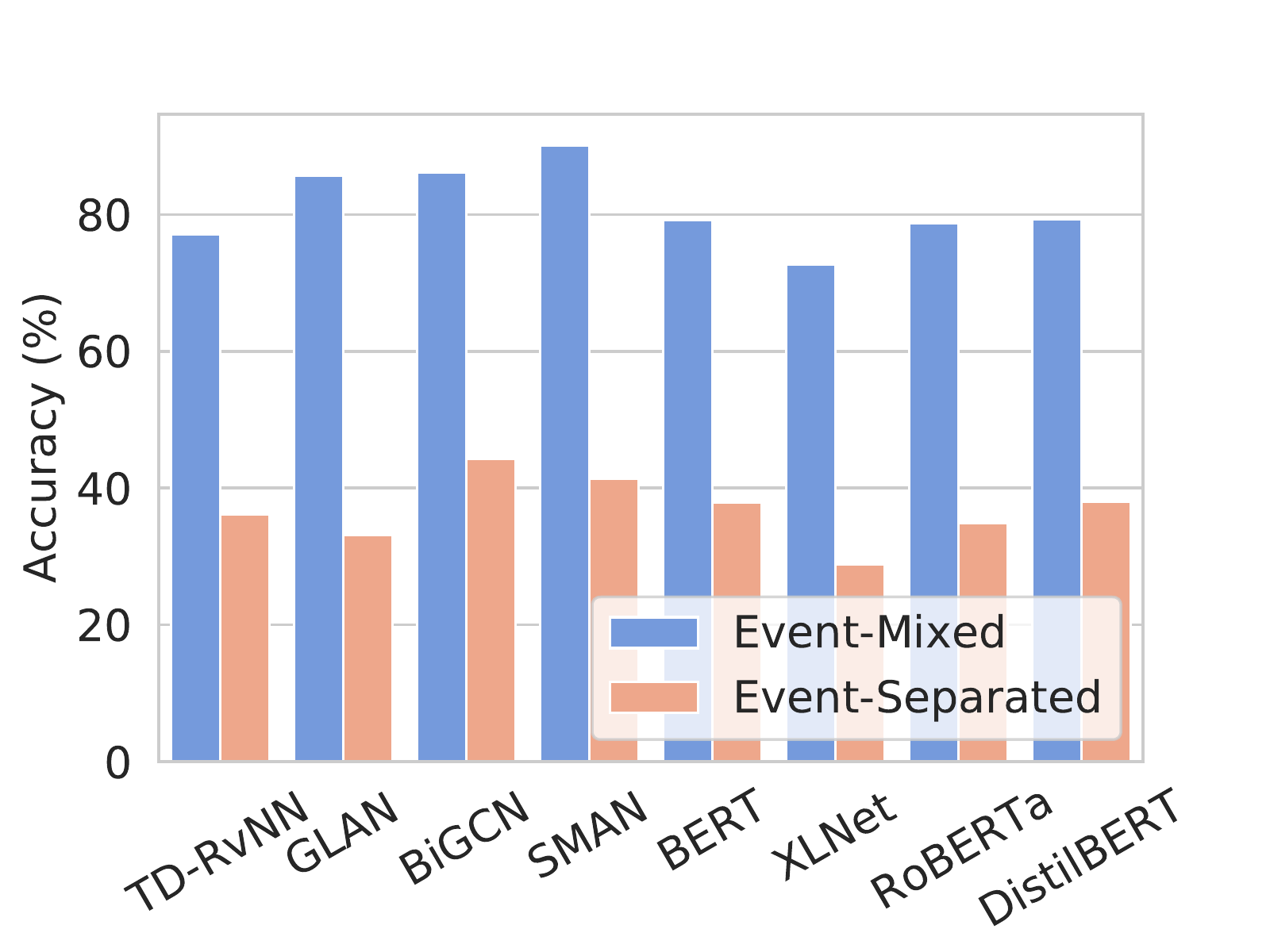}}
    \subfigure[\textsc{PHEME}]{\includegraphics[width=0.32\textwidth]{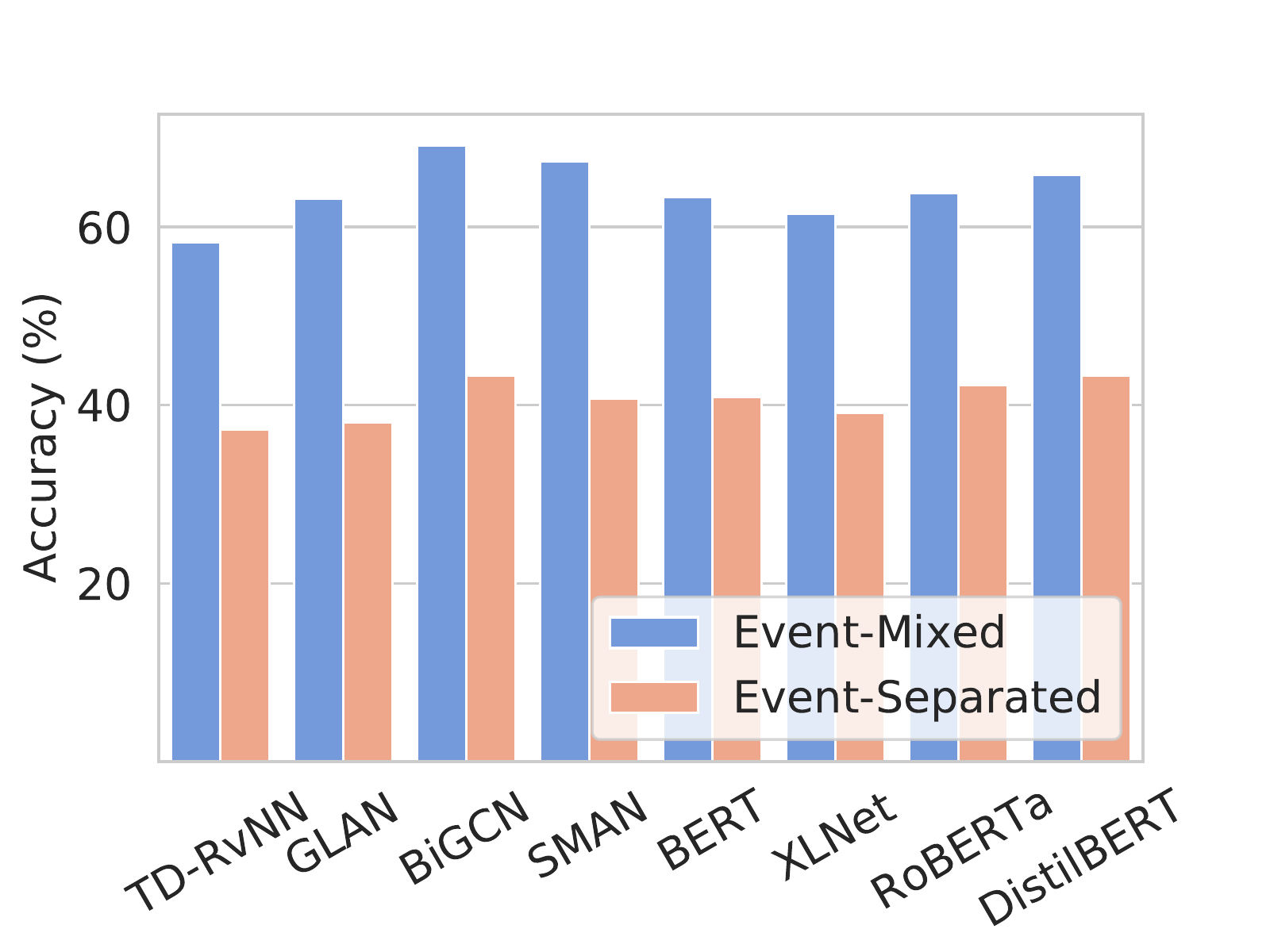}} 
    \caption{\textbf{Existing rumor detection approaches fail to generalize across events.} Comparing between event-mixed and event-separated settings, mean accuracy based on 20 different runs of each approach demonstrates drastic performance deterioration. (Over 40\% drop on \textsc{Twitter15} \& \textsc{16}, and over 20\%  on \textsc{PHEME}.)}
    \label{fig:model-drop}
\end{figure*}

\subsection{SOTA Models' Performance is Heavily Overestimated} 

Fig. \ref{fig:model-drop} reveals a stark contrast between event-mixed and event-separated rumor detection performance. More specifically, test accuracy plummets from 74.0\%$\sim$89.2\% to 30.5\%$\sim$42.8\% on \textsc{Twitter15}, from 72.7\%$\sim$90.2\% to 28.8\%$\sim$44.3\% on \textsc{Twitter16}, and from 58.3\%$\sim$69.2\% to 37.3\%$\sim$43.4\% on \textsc{PHEME}.
Furthermore, despite the consistency of best event-separated performance across all three datasets, all models achieve significantly higher event-mixed performance on \textsc{Twitter15} and \textsc{Twitter16} than on \textsc{PHEME}, where the former adopts event-level labeling and the latter adopts post-level labeling (see Section \ref{data-collect}). This gap is in line with our hypothesis that direct event-label correlations induce additional bias.

Results imply the heavy reliance of existing methods on spurious event-specific correlations. Despite performing well under the event-mixed setting, these models cannot generalize to unseen events, resulting in poor real-world adaptability.

\section{Proposed Method}

\begin{figure}[t] 
 \centerline{\includegraphics[width=0.9\textwidth]{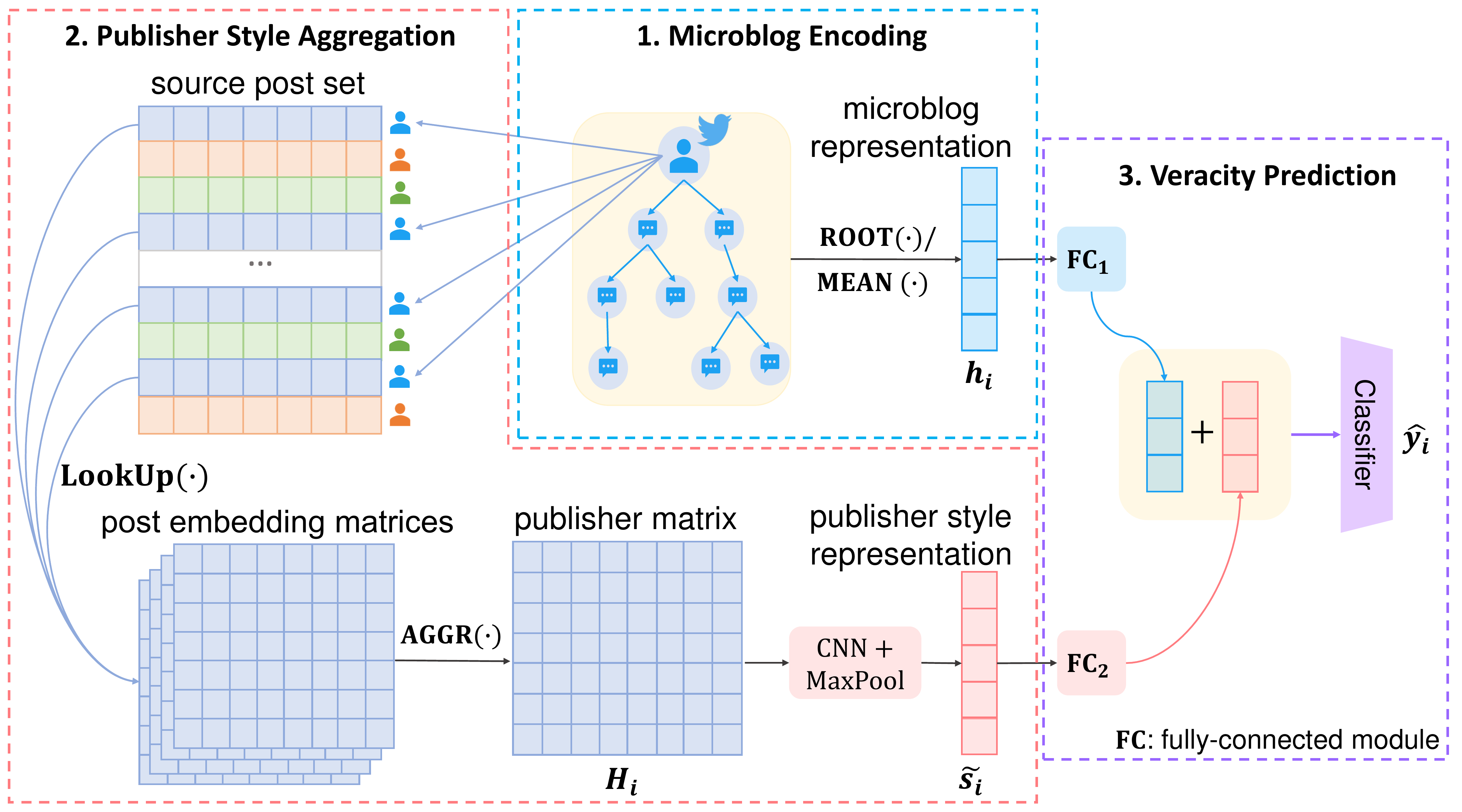}}
 \caption{Overview of our proposed PSA framework.}
 \label{fig:proposed-framework} 
\end{figure}

To tackle the challenges of event-separated rumor detection, we propose Publisher Style Aggregation (PSA), a novel approach that learns generalizable publisher characteristics based on each publisher's aggregated posts, as illustrated in Fig.\ref{fig:proposed-framework}.

\begin{figure*}[t]
    \centering
    \subfigure[Non-Rumor]{\includegraphics[width=0.24\textwidth]{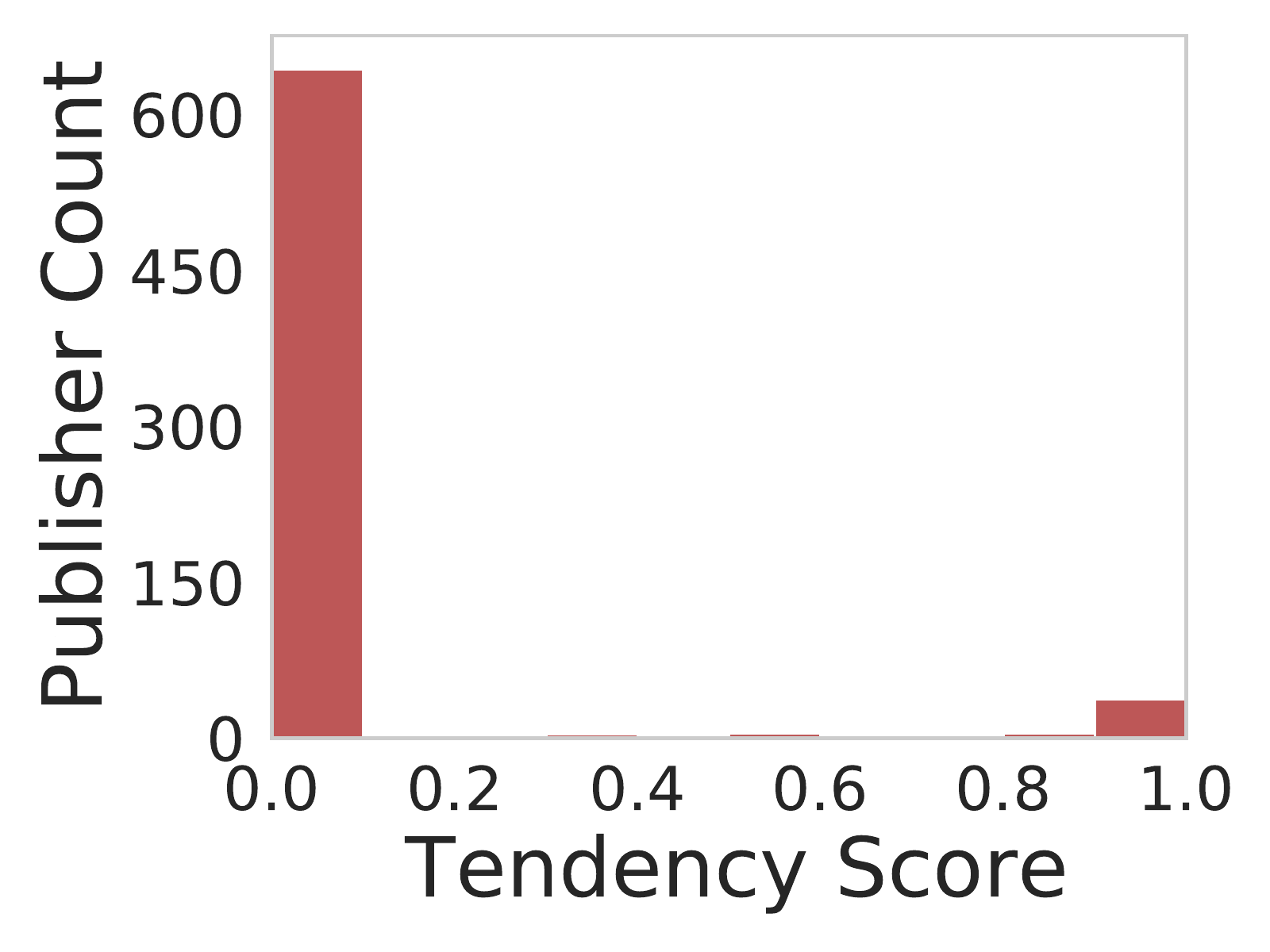}} 
    \subfigure[False]{\includegraphics[width=0.24\textwidth]{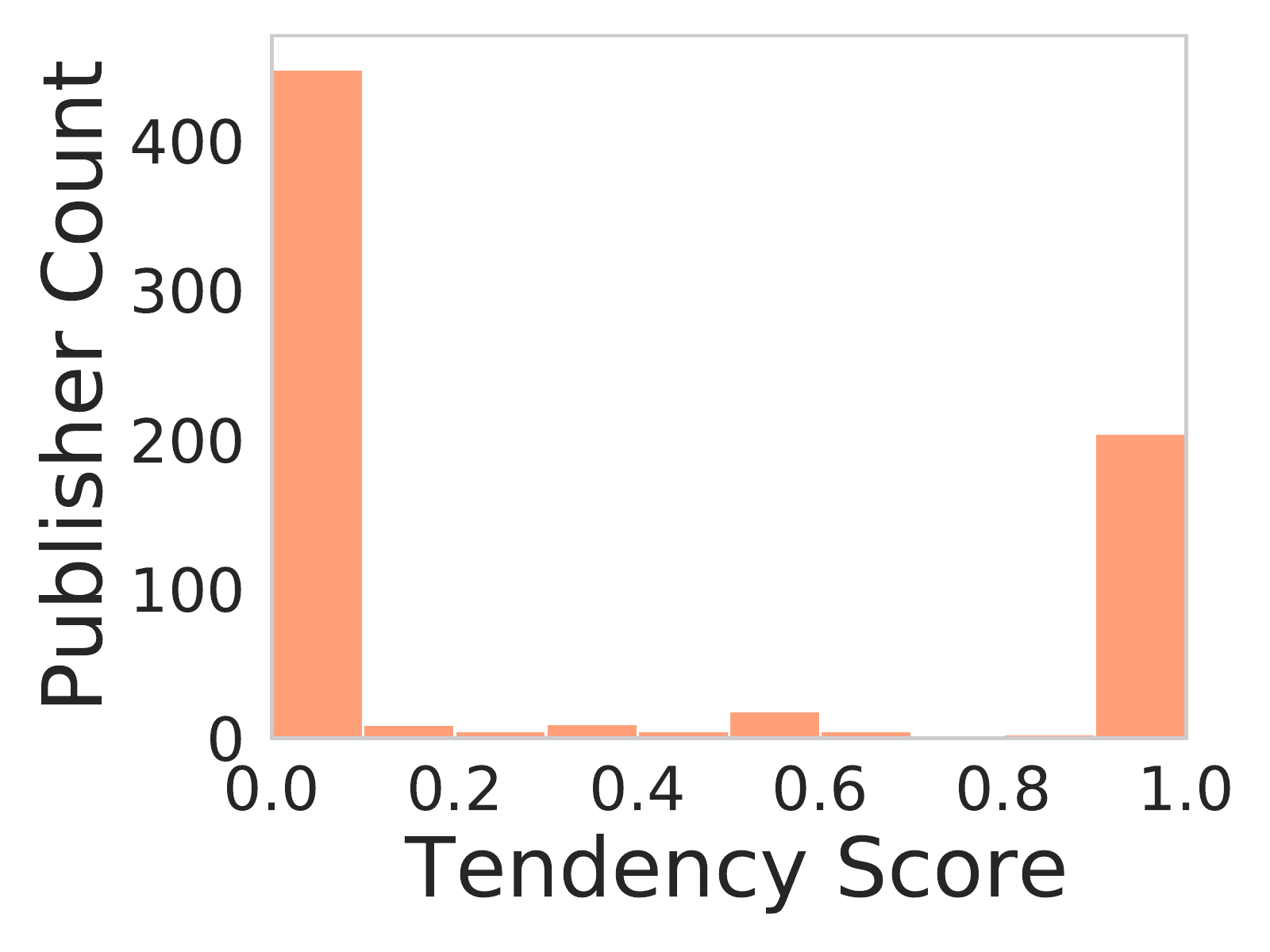}} 
    \subfigure[True]{\includegraphics[width=0.24\textwidth]{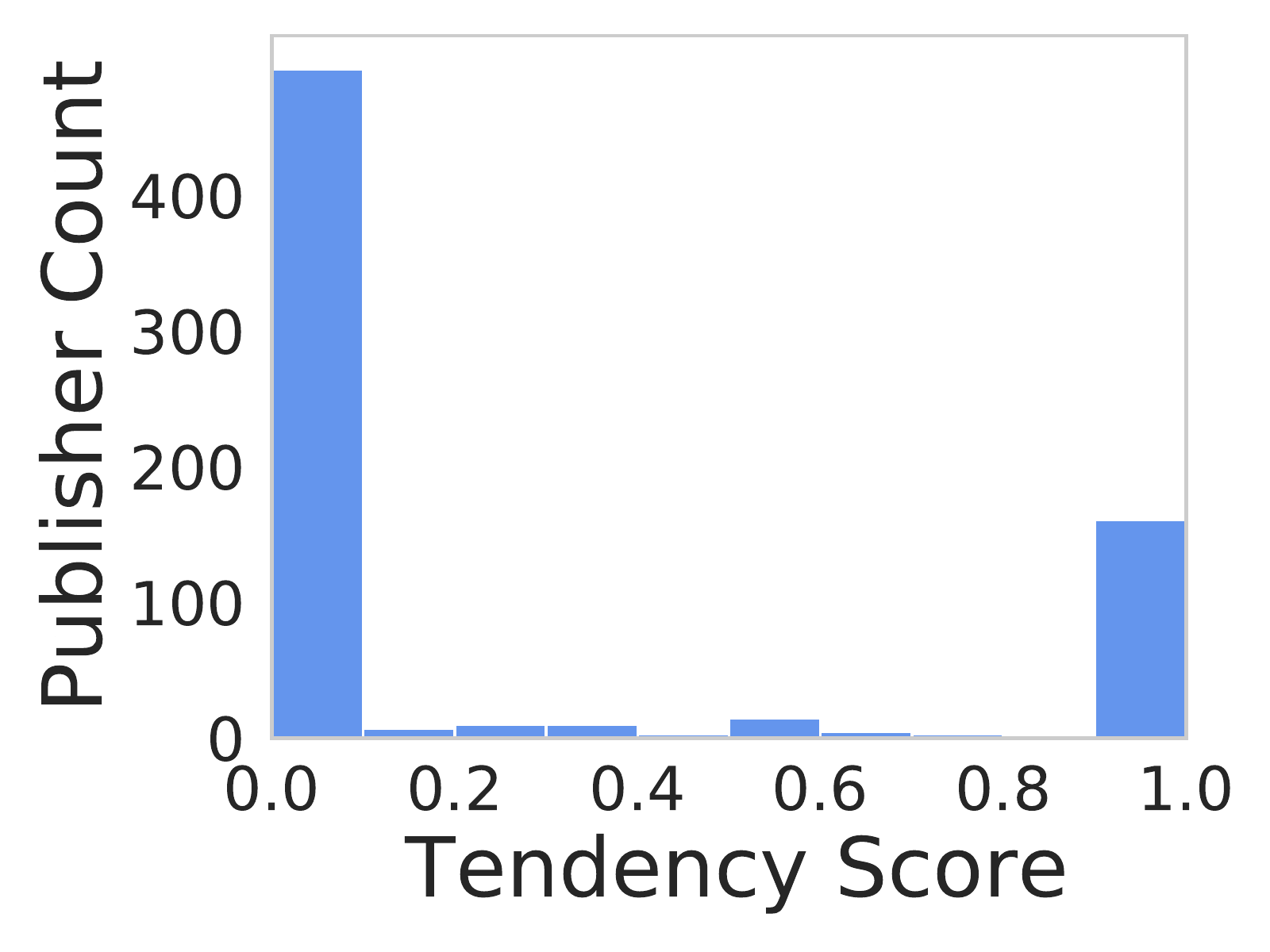}} 
    \subfigure[Unverified]{\includegraphics[width=0.24\textwidth]{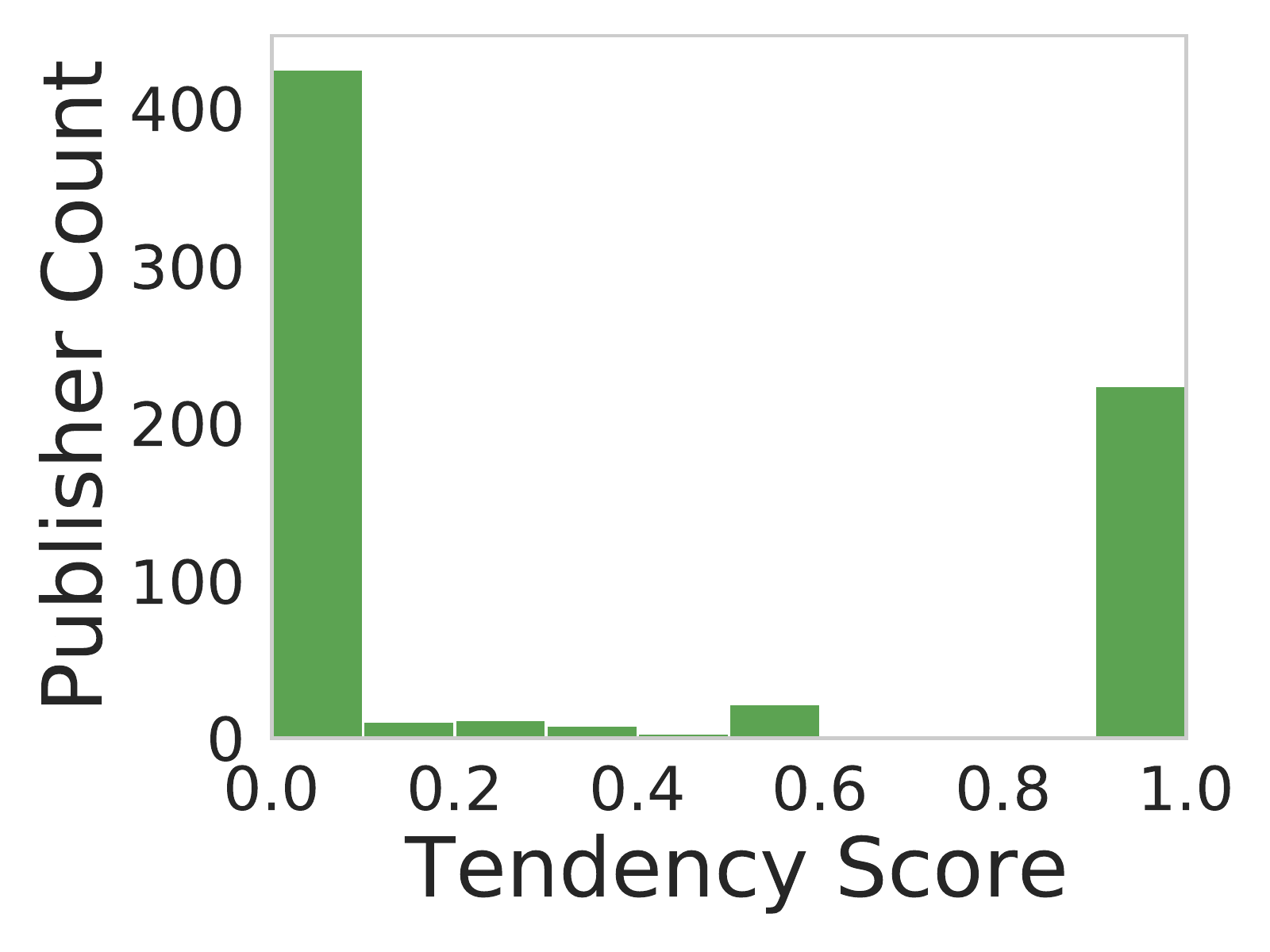}} 
    \caption{\textbf{Publishers tend to post tweets of the same credibility type}, as exemplified by \textsc{Twitter15} publisher behavior patterns.}
    \label{fig:pub_style}
\end{figure*}

\subsection{Consistency of Publisher Style} 
\label{sec:cons_pub}

Source post publishers are highly influential users who produce claims towards newsworthy events. Therefore, each publisher's unique credibility stance and writing style can exhibit distinctive traits that help determine the veracity of their statements. For a more intuitive view, we illustrate the \textsc{Twitter15} publisher tendency towards each class in Fig. \ref{fig:pub_style}. Specifically, for publisher $u$, we define $u$'s tendency score under class $c$ as (\# microblogs posted by $u$ under class $c$)/(\# microblogs posted by $u$). Fig. \ref{fig:pub_style} shows that most publishers have scores either approaching 0 or approaching 1 towards one particular class, i.e., most publishers tend to post microblogs under a single veracity label, which verifies our hypothesis of publisher style consistency.

\subsection{Content-Based Microblog Encoding} 
We first propose two simple neural classifiers, namely RootText and MeanText, to study the importance of source post and comment features in social media rumor detection. 
In each dataset, all source posts and comments constitute a vocabulary of size $|V| $. Following \cite{bian20rumor,ma17detect}, we formulate each source post feature $\mathbf{r}_{i}\in \mathbb{R}^{|V|}$ and its related comment features $\mathbf{r}_{i}^{j}\in \mathbb{R}^{|V|}$ as the sum of all one-hot word vectors within the corresponding source post or comment.

\underline{\textbf{RootText:}} Source posts contain the claims to be fact-checked. Therefore, we propose to encode each microblog instance $T_{i}$ solely based on its source post textual feature $\mathbf{r}_{i}$, i.e.,  $\mathbf{h}_{i}:=\mathbf{r}_{i}$.

\underline{\textbf{MeanText:}} We also propose to consider the user comments of source posts for more robust credibility measurement. Here, we adopt mean-pooling to condense source post and comment features into microblog representations:
\begin{equation}
    \mathbf{h}_{i}:=\frac{\mathbf{r}_{i}+{\scriptstyle\sum}_{j=1}^{k} \mathbf{r}_{i}^{j}}{k+1}. \label{eq:meantext}
\end{equation} 

We obtain the encoding $\mathbf{h}_i \in \mathbb{R}^{
|V|}$ of microblog $T_i$ based on either RootText or MeanText, and extract high-level features $\tilde{\mathbf{h}}_i \in \mathbb{R}^{n}$ via a two-layer fully-connected neural network with the ReLU activation function. 
Then, we employ dropout to prevent overfitting before passing $\tilde{\mathbf{h}}_i$ through the final fully-connected layer with output dimensionality $|\mathcal{C}|$ for veracity prediction.

\subsection{Publisher Style Aggregation} 
As shown in Section \ref{sec:cons_pub}, the writing stance and credibility of highly influential source post publishers remain relatively stable in a fixed timeframe. Inspired by this observation, we further propose Publisher Style Aggregation (PSA), a generalizable method that jointly leverages multiple microblog instances produced by each publisher and extracts distinctive publisher traits to enhance local features learned within each individual microblog. More specifically, PSA (1) looks up a set of microblog instances produced by each publisher, (2) learns publisher style representations based on these source posts' aggregated textual features, and (3) augments the representation of each microblog (i.e. $\tilde{\mathbf{h}}_i$ learned via RootText / MeanText) with its corresponding publisher style representation $\tilde{\mathbf{s}}_i$. 

\textbf{Publisher Style Modeling:} Assume that publisher ${u}_i$ has produced $m_i\ge 1$ microblog instances, with the corresponding source posts denoted as $\mathcal{P}(u_i)=\left\{p_k|u_k=u_i,k=1,\dots,N\right\}$; note that only accessible data are used during training. We treat the $j$-th post $p_{i}^{j}\in \mathcal{P}(u_i)$ as a word token sequence with maximum length $L$. Then, we construct an embedding matrix $\mathbf{W}_{i}^{j} \in \mathbb{R}^{L\times d}$ for $p_{i}^{j}$ based on trainable $d$-dimensional word embeddings. We aggregate all post embedding matrices $\{\mathbf{W}_{i}^{j}\}_{j=1}^{m_i}$ of $u_{i}$, and obtain the corresponding publisher matrix $\mathbf{H}_{i}\in \mathbb{R}^{L\times d}$ as follows:
\begin{equation}
    \mathbf{H}_{i}=\mathsf{AGGR}(\{\mathbf{W}_{i}^{j}\}_{j=1}^{m_i}), 
    \label{eq:hist-aggr}
\end{equation}
where the $\mathsf{AGGR}$ operator can be either $\mathsf{MEAN}$ or $\mathsf{SUM}$. 

To capture high-level publisher characteristics, we apply convolution on each $\mathbf{H}_{i}$ to extract latent publisher style features. Specifically, we use three convolutional layers with different window sizes to learn features with varied granularity. Each layer consists of $F$ filters, and each filter outputs a feature map $\mathbf{f_{*}}=[f_{*}^{1},f_{*}^{2},\ldots,f_{*}^{L-k+1}]$, with
\begin{equation}
   f_{*}^{j}=\mathsf{ReLU}\left(\mathbf{W}_{f} \cdot \mathbf{H}_{i}[j:j+k-1]+b\right),
\end{equation}
where $\mathbf{W}_{f} \in \mathbb{R}^{k\times d}$ the convolution kernel, $k$ the window size and $b \in \mathbb{R}$ a bias term. We perform max-pooling to extract the most prominent value of each $\mathbf{f}_{*}$, and stack these values to form a style feature vector $\mathbf{s} \in \mathbb{R}^{F}$. Then, we concatenate the $\mathbf{s}_{*}$ produced by each of the three CNN layers to obtain the publisher style representation $\tilde{\mathbf{s}}_{i} \in \mathbb{R}^{3F}$:
\begin{equation}
    \tilde{\mathbf{s}}_{i}=\mathsf{Concat}[\mathbf{s}_{1}; \mathbf{s}_{2}; \mathbf{s}_{3}].
\end{equation}

\textbf{Microblog Veracity Prediction:} We augment microblog representation $\tilde{\mathbf{h}}_i\in \mathbb{R}^{n}$ with the corresponding publisher style representation $\tilde{\mathbf{s}}_{i}$. Finally, we utilize a fully connected layer to predict the microblog veracity label $\hat{\mathbf{y}}_i$:
\begin{equation}
    \hat{\mathbf{y}}_{i}=\mathsf{Softmax}( \mathbf{W}_2^\mathsf{T}(\tilde{\mathbf{h}}_i + \mathbf{W}_1^\mathsf{T}\tilde{\mathbf{s}}_i)),
\end{equation}
where transformations $\mathbf{W}_{1} \in \mathbb{R}^{3F\times n}$ and $\mathbf{W}_{2} \in \mathbb{R}^{n\times |\mathcal{C}|}$. We also apply dropout before the final layer to prevent overfitting. 

Model parameters are optimized by minimising the cross-entropy loss between $\hat{\mathbf{y}}_i$ and ground truth $y_i$.

\section{Experiments} \label{experiments}

In this section, we review our experiments for answering the following questions:

\textbf{Q1 (Model Performance):} Does PSA outperform the existing baselines on event-separated rumor detection? 

\textbf{Q2 (Early Rumor Detection):} Does PSA work well under temporal rumor detection deadlines?

\textbf{Q3 (Model Generalization):} Is PSA effective under cross-dataset settings?

\subsection{Experimental Setup}
We implement our proposed PSA model and its variants based on PyTorch 1.6.0 with CUDA 10.2, and train them on a server running Ubuntu 18.04 with NVIDIA RTX 2080Ti GPU and Intel(R) Xeon(R) CPU E5-2690 v4 @ 2.60GHz. We adopt an Adam optimizer with $(\beta_1, \beta_2)=(0.9, 0.999)$, learning rate of $10^{-4}$ ($0.005$), and weight decay $10^{-5}$ ($10^{-4}$) for \textsc{Twitter15/16} (\textsc{PHEME}). We obtain microblog representations via a $2$-layer neural network with layer sizes of $128$ and $64$. We utilize the $300$-dimensional word vectors from \cite{yuan19jointly} to form publisher matrices, employ three CNN layers with the same filter number $F=100$ but different window sizes $k \in\{3,4,5\}$, concatenate their outputs and use a fully-connected layer to extract publisher style representations with the size of $64$. We implement $\mathsf{AGGR}$ in Eq. \ref{eq:hist-aggr} as both $\mathsf{SUM}$ and $\mathsf{MEAN}$, and report average performance over 20 different runs on the event-separated data splits presented in Section \ref{existing-models}.

\subsection{Q1. Model Performance} \label{model-performance}

We compare PSA (base classifier: RootText / MeanText) with existing approaches
in Table \ref{tab:psa}.

\begin{table*}[t]
\centering
\caption{PSA significantly improves event-separated rumor detection accuracy (\%) and Macro F1 Score (\%) ($\mathsf{S}$ stands for $\mathsf{SUM}$ and $\mathsf{M}$ for $\mathsf{MEAN}$; averaged over 20 runs).}\label{tab:psa}
\begin{tabular}{lcccccccc}\toprule
\multirow{2}{*}{\textbf{Method}} & \multicolumn{2}{c}{\textsc{Twitter15}} & & \multicolumn{2}{c}{\textsc{Twitter16}} & &\multicolumn{2}{c}{\textsc{PHEME}} \\
\cmidrule{2-3} \cmidrule{5-6} \cmidrule{8-9}
& Acc. & F1  & &Acc. & F1 & &Acc.  & F1 \\
\midrule
TD-RvNN \cite{ma18rumor}& 38.62\textsubscript{$\pm$1.85} & 36.40\textsubscript{$\pm$2.38} && 36.15\textsubscript{$\pm$1.90} & 35.66\textsubscript{$\pm$1.89} && 37.30\textsubscript{$\pm$2.54} &34.17\textsubscript{$\pm$2.56}\\
GLAN \cite{yuan19jointly} & 38.56\textsubscript{$\pm$3.38} & 35.52\textsubscript{$\pm$5.31} && 33.13\textsubscript{$\pm$4.54} & 27.93\textsubscript{$\pm$5.53} && 38.10\textsubscript{$\pm$2.85} & 34.60\textsubscript{$\pm$3.04} \\
BiGCN \cite{bian20rumor} & 42.83\textsubscript{$\pm$2.27} & 38.17\textsubscript{$\pm$3.04} && 44.28\textsubscript{$\pm$3.39} & 42.31\textsubscript{$\pm$3.77} && 43.36\textsubscript{$\pm$1.71} & 37.93\textsubscript{$\pm$2.16}\\
SMAN \cite{yuan20early}& 30.52\textsubscript{$\pm$2.62} & 28.80\textsubscript{$\pm$3.30} && 41.42\textsubscript{$\pm$2.65} & 40.62\textsubscript{$\pm$2.95} && 40.74\textsubscript{$\pm$1.36} & 36.02\textsubscript{$\pm$1.62} \\
BERT \cite{devlin19bert} & 40.95\textsubscript{$\pm$4.80}  & 37.47\textsubscript{$\pm$7.56} && 37.89\textsubscript{$\pm$6.68}  & 34.76\textsubscript{$\pm$9.22} && 40.90\textsubscript{$\pm$3.22} & 36.33\textsubscript{$\pm$4.02} \\
XLNet \cite{yang19xlnet} & 32.05\textsubscript{$\pm$6.78} & 26.00\textsubscript{$\pm$9.20} && 28.82\textsubscript{$\pm$4.08} & 20.59\textsubscript{$\pm$7.06} && 39.14\textsubscript{$\pm$5.14} & 34.35\textsubscript{$\pm$6.65} \\
RoBERTa \cite{liu19roberta} & 35.30\textsubscript{$\pm$5.65} & 28.41\textsubscript{$\pm$7.93} && 34.84\textsubscript{$\pm$6.69} & 27.90\textsubscript{$\pm$9.74} && 42.25\textsubscript{$\pm$4.47} &  37.67\textsubscript{$\pm$5.70} \\
DistilBERT \cite{sanh20distilbert} & 38.09\textsubscript{$\pm$4.48} & 33.74\textsubscript{$\pm$5.21} && 38.02\textsubscript{$\pm$4.24} & 33.98\textsubscript{$\pm$5.56} && 43.33\textsubscript{$\pm$3.62} & 38.37\textsubscript{$\pm$4.16}\\
\midrule
RootText(RT) & 33.80\textsubscript{$\pm$2.74} & 30.56\textsubscript{$\pm$2.92} && 30.54\textsubscript{$\pm$1.72} & 28.87\textsubscript{$\pm$2.32} && 42.75\textsubscript{$\pm$1.25} & 38.74\textsubscript{$\pm$1.47}\\
MeanText(MT) &  48.68\textsubscript{$\pm$1.80}  & 47.18\textsubscript{$\pm$1.63} && 45.19\textsubscript{$\pm$1.86}  & 44.23\textsubscript{$\pm$1.72} && 32.57\textsubscript{$\pm$1.70} & 30.48\textsubscript{$\pm$1.51} \\
\midrule
RT+PSA($\mathsf{S}$) & 47.85\textsubscript{$\pm$5.64} & 45.26\textsubscript{$\pm$5.13} && 57.88\textsubscript{$\pm$3.16} & 55.30\textsubscript{$\pm$4.64} && 43.52\textsubscript{$\pm$0.93} & 37.81\textsubscript{$\pm$1.22} \\
RT+PSA($\mathsf{M}$) & 45.67\textsubscript{$\pm$0.82} & 38.55\textsubscript{$\pm$0.89} && 47.28\textsubscript{$\pm$2.87} & 42.74\textsubscript{$\pm$4.15} &&  \textbf{46.30}\textsubscript{$\pm$1.28} & \textbf{41.57}\textsubscript{$\pm$1.49} \\

MT+PSA($\mathsf{S}$) & \textbf{61.83}\textsubscript{$\pm$1.43} & \textbf{58.75}\textsubscript{$\pm$2.08} &&\textbf{64.89}\textsubscript{$\pm$1.75} & \textbf{64.31}\textsubscript{$\pm$1.70} && 37.43\textsubscript{$\pm$1.05} & 32.36\textsubscript{$\pm$1.25} \\
MT+PSA($\mathsf{M}$) & 54.73\textsubscript{$\pm$1.04} & 50.85\textsubscript{$\pm$1.56} && 60.16\textsubscript{$\pm$2.76} & 58.16\textsubscript{$\pm$3.13} && 40.19\textsubscript{$\pm$1.20} &  35.98\textsubscript{$\pm$1.45} \\
\bottomrule
\end{tabular}
\end{table*}

\textbf{Importance of Textual Features:} We observe that MeanText outperforms existing methods on \textsc{Twitter15\&16}, while RootText only achieves 0.6\% lower accuracy than the best baseline on \textsc{PHEME}. This implies severe degradation of overparameterized models when the spurious attribute-label correlations (i.e. event-specific cues) in the training data do not apply to the test data, in line with prior work \cite{sagawa20inv}. Comparing between RootText and MeanText, we also observe that the former performs better on \textsc{PHEME} but otherwise on \textsc{Twitter15} and \textsc{Twitter16}. Different labeling schemes may account for such differences; as \textsc{PHEME} labels each microblog independently, the source posts would contain the most distinctive features. While the source post content is not as distinctive in \textsc{Twitter15} and \textsc{Twitter16}, both datasets exhibit more complicated propagation patterns (see Table \ref{tab:ds-stats}). Therefore, adopting MeanText to aggregate comment features proves more effective in these cases.

\textbf{Effectiveness of PSA:} Our proposed PSA approach, with $\mathsf{AGGR}$ implemented as either $\mathsf{SUM}$ or $\mathsf{MEAN}$, significantly enhances the base classifiers RootText and MeanText. The best PSA combinations outperform the best baseline by a large margin; they boost event-separated rumor detection accuracy by 19.00\% on \textsc{Twitter15}, 20.61\% on \textsc{Twitter16}, and 2.94\% on \textsc{PHEME}. Unlike existing methods, PSA explicitly aggregates publisher style features across microblog instances from multiple events, thereby enhancing the model's capability to learn event-invariant features. As a result, PSA is able to capture stance and style pertaining to distinctive publisher characteristics, leading to substantial performance improvements. 

\begin{figure*}[t]
    \centering
    \subfigure[\textsc{Twitter15}]{\includegraphics[width=0.32\textwidth]{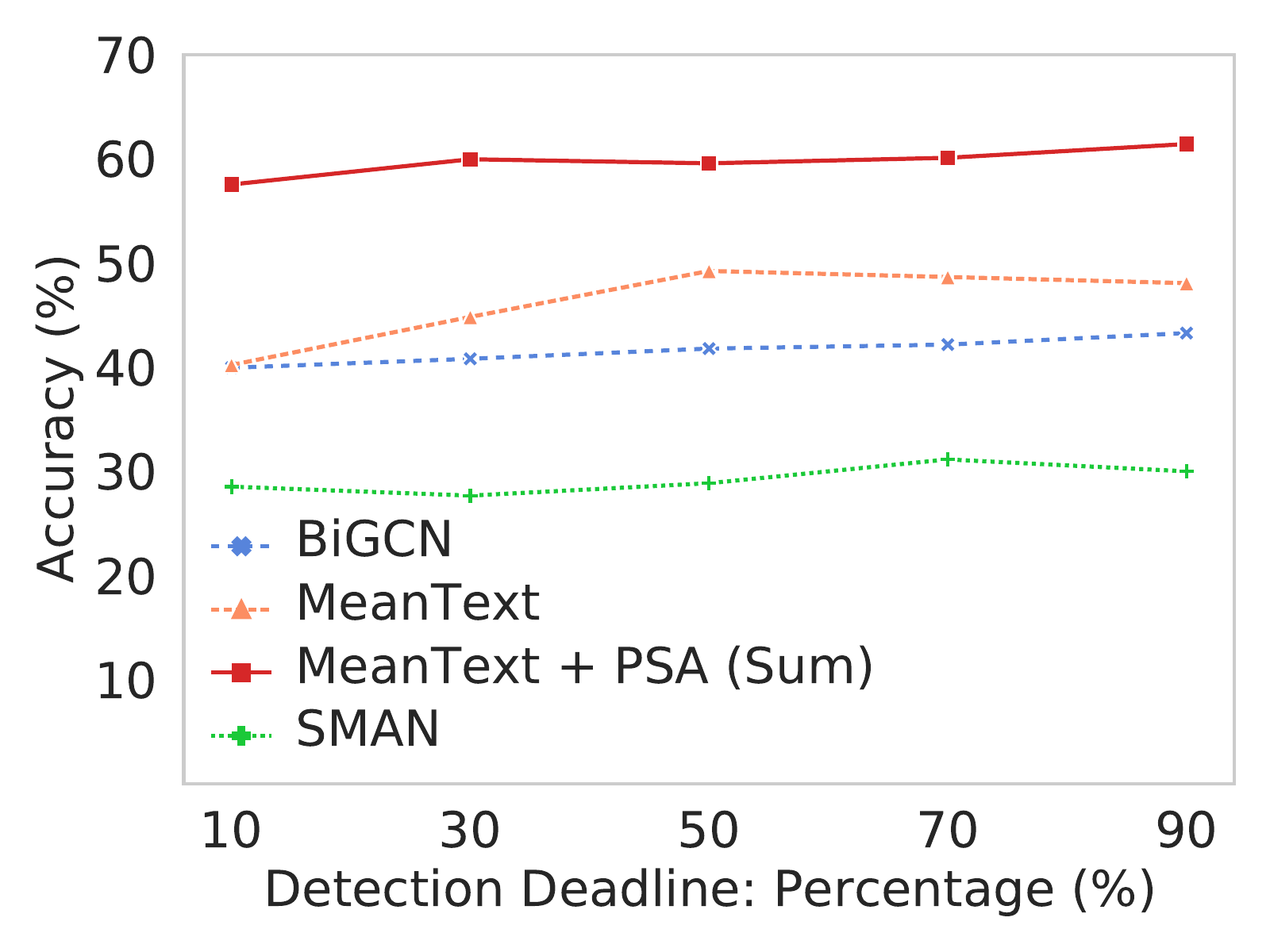}} 
    \subfigure[\textsc{Twitter16}]{\includegraphics[width=0.32\textwidth]{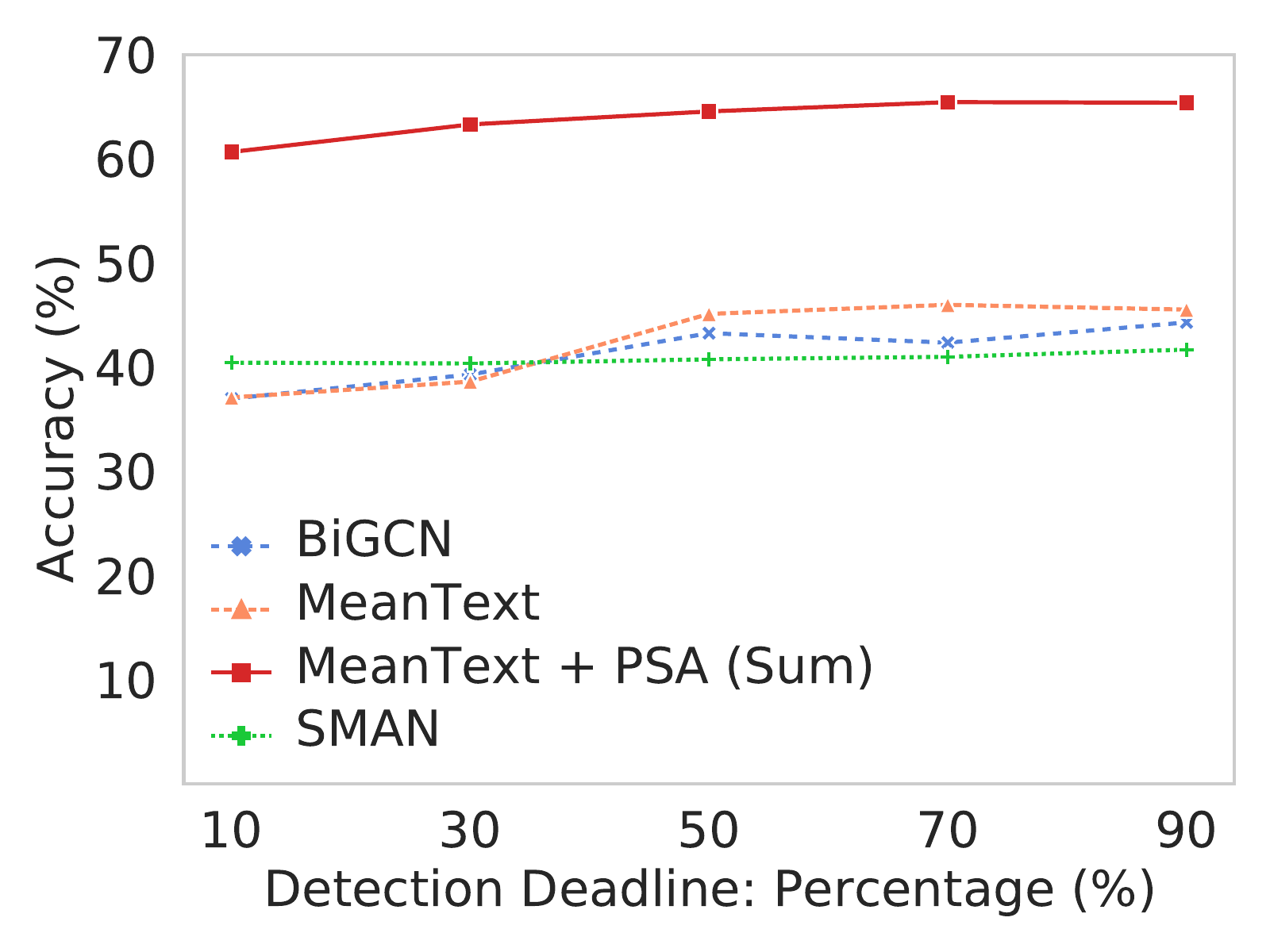}} 
    \subfigure[\textsc{PHEME}]{\includegraphics[width=0.32\textwidth]{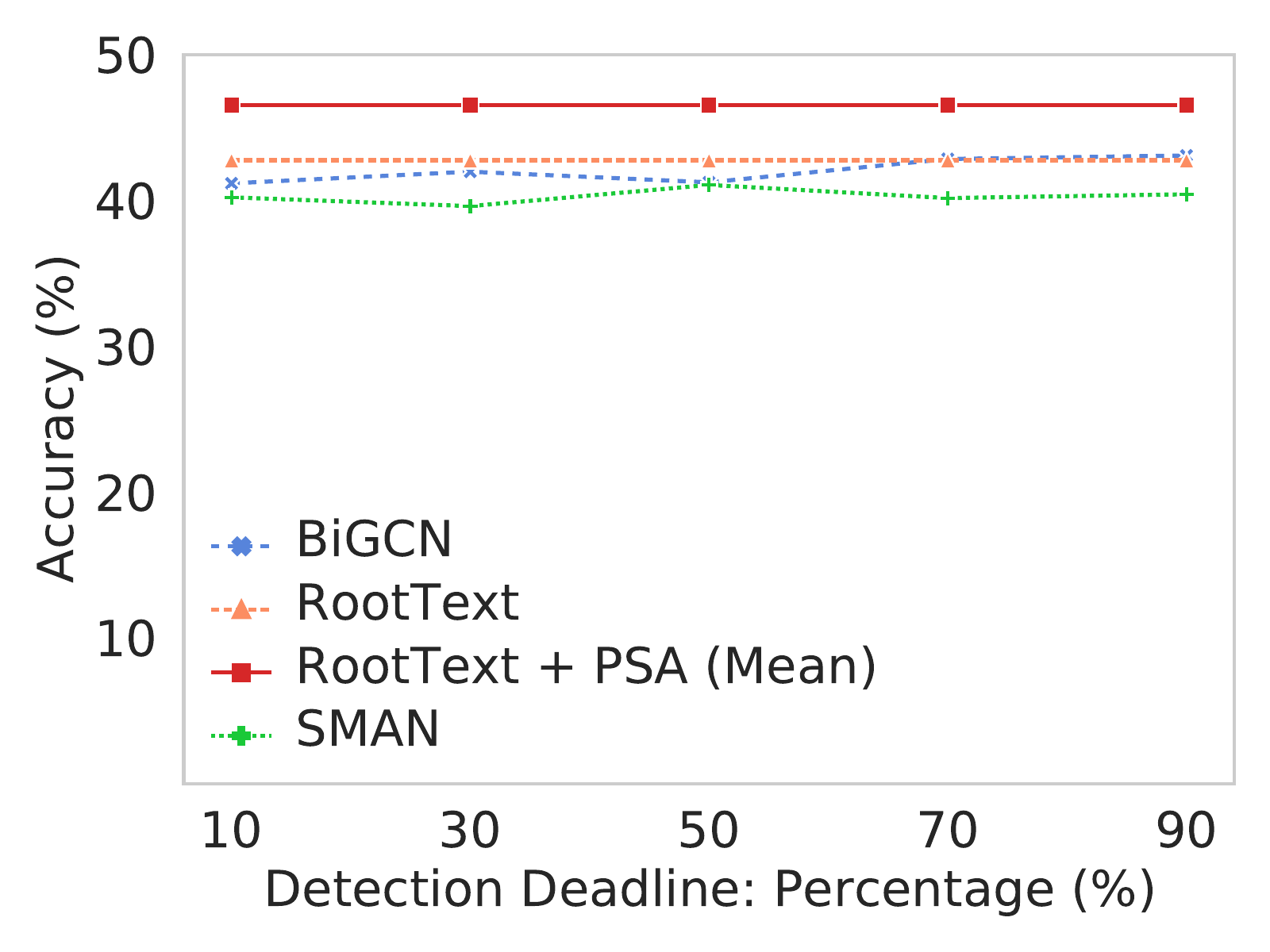}} 
    \caption{Early rumor detection accuracy (\%) of PSA models against the best propagation-based baselines for varying detection deadlines on \textsc{Twitter15}, \textsc{Twitter16} and \textsc{PHEME} under the event-separated setting (averaged over 20 runs).}
    \label{fig:early-detect}
\end{figure*}

\subsection{Q2. Early Rumor Detection} \label{early-detect}

Accurate and timely misinformation detection is of vital importance. Given only partial propagation information, we compare the best PSA combinations and their corresponding base classifiers with their best propagation-based competitors. Fig. \ref{fig:early-detect} shows the consistent superiority of PSA over baseline methods at all detection deadlines. Even with only the earliest 10\% of comments, PSA achieves 57.53\% accuracy on \textsc{Twitter15}, 60.65\% on \textsc{Twitter16}, and 46.30\% on \textsc{PHEME}. Note that RootText (+PSA) models maintain stable performance across all deadlines, as they provide instant predictions solely based on source posts. The results demonstrate that augmenting rumor detection models with publisher style representations achieves both efficiency and effectiveness.

\begin{table}[t]
\centering 
\caption{Cross-dataset rumor detection accuracy (\%) and Macro F1 Score (\%) of PSA between \textsc{T15} (\textsc{Twitter15}) and \textsc{T16} (\textsc{Twitter16}), compared with the best propagation- and content-based baseline methods (averaged over 20 runs).}\label{tab:cross-dset}
 \begin{tabular}{cclcc} \toprule
 Train & Test & \textbf{Method} & Acc. & F1 \\
 \toprule
 \multirow{4}{*}{\textsc{T15}} & \multirow{4}{*}{\textsc{T16}} & 
 DistilBERT & 40.70\textsubscript{$\pm$4.78} & 39.98\textsubscript{$\pm$5.02}  \\ & &
 BiGCN & 36.80\textsubscript{$\pm$3.79} & 35.76\textsubscript{$\pm$3.92}  \\ & &
 MeanText & 46.55\textsubscript{$\pm$1.33} & 44.11\textsubscript{$\pm$1.71} \\& &
 + PSA (Sum) & \textbf{63.99}\textsubscript{$\pm$1.53} & \textbf{61.95}\textsubscript{$\pm$1.86} \\ 
 \midrule
 \multirow{4}{*}{\textsc{T16}} & \multirow{4}{*}{\textsc{T15}} & 
 DistilBERT & 41.66\textsubscript{$\pm$ 4.95} & 37.07\textsubscript{$\pm$6.06}  \\ & &
 BiGCN & 44.39 \textsubscript{$\pm$1.89} & 42.41\textsubscript{$\pm$2.23}  \\ & &
 MeanText & 48.00 \textsubscript{$\pm$1.70} & 44.83\textsubscript{$\pm$1.63}  \\& &
 + PSA (Sum) & \textbf{60.82}\textsubscript{$\pm$1.47}  & \textbf{57.97}\textsubscript{$\pm$2.07} \\ 
 \bottomrule
\end{tabular} 
\end{table}
\subsection{Q3. Cross-Dataset Rumor Detection} 

To study the generalization ability of PSA, we conduct cross-dataset experiments on \textsc{Twitter15} and \textsc{Twitter16}, where the model is trained on one dataset and tested on the other. For a fair comparison, we utilize the same event-separated data splits adopted in Sections \ref{model-performance} and \ref{early-detect}. If overlapping events exist between the training set from dataset A and the test set from dataset B, we remove all instances related to these events in the training set, and replace them with the same number of non-overlapping instances randomly sampled from 
A's test set. 

The cross-dataset setting is inherently more challenging, as the training and test events stem from different timeframes, which can create temporal concept shifts. However, Table \ref{tab:cross-dset} shows that PSA continues to excel and enhances the base classifier (our MeanText method) by 17.44\% on \textsc{Twitter15} and 12.82\% on \textsc{Twitter16}, which further demonstrates PSA's generality to unseen events.

\subsection{Discussion: Source-Specific Spurious Cues} \label{event-insensitive}

In Table \ref{tab:insensitive-cues}, we empirically show the potential impact of source-specific spurious cues (Section \ref{spurious-cues}). Under the event-separated setting, we construct a simple Random Forest classifier based on each microblog's user interaction count and the time range covered by these interactions. Surprisingly, the classifier outperforms existing baseline methods on both \textsc{Twitter15} and \textsc{Twitter16}, and achieves comparable performance even with only one feature. In contrast, the single-source \textsc{PHEME} remains unaffected. Although neither our proposed approaches nor existing methods exploit these features, we nevertheless suggest the integration of debiasing techniques in future graph- and temporal-based models.

\begin{table}[t]
\centering 
\caption{Classification accuracy (\%) based on user interaction count and interaction time range suggests potential source-specific correlations.}\label{tab:insensitive-cues}
 \begin{tabular}{lccc} \toprule
  \textbf{Feature}& \textsc{Twitter15} & \textsc{Twitter16} & \textsc{PHEME}\\
 \toprule
 A: Interaction Count & 43.03 & 43.32 & 23.15\\
 B: Time Range & 37.09 & 44.92 & 27.31\\ 
 A + B & \textbf{44.21} & \textbf{55.61} & 26.85 \\
 \bottomrule
\end{tabular} 
\end{table}

\section{Conclusion} 

In this paper, we systematically analyze how event-based data collection schemes create event- and source-specific spurious correlations in social media rumor detection benchmark datasets. We study the task of event-separated rumor detection to remove event-specific correlations, and empirically demonstrate severe limitations on existing methods' generalization ability. To better address this task, we propose PSA to augment microblog representations with aggregated publisher style features. Extensive experiments on three real-world datasets show substantial improvement on cross-event, cross-dataset and early rumor detection.

For future work, we suggest (1) event-separated rumor detection performance as a major evaluation metric; (2) same-source samples and post-level expert annotations in dataset construction; and (3) integration of causal reasoning and robust learning techniques in model design, in the hope that our findings could motivate and measure further progress in this field.

\subsubsection{Acknowledgements} 
This work was supported in part by NUS ODPRT Grant R252-000-A81-133.

%
%
%
%

\end{document}